\documentstyle[prb,aps]{revtex}
\tightenlines
\begin{document}
\draft
\input{psfig}
\title{
Variable and reversible quantum structures on a single carbon nanotube
}
\author{\c{C}. K{\i}l{\i}\c{c}$^{(1)}$, S. Ciraci$^{(1)}$, 
O. G\"{u}lseren$^{(2,3)}$, and T. Yildirim$^{(2)}$}
\address{$^{(1)}$ Physics Department, Bilkent University, Ankara, Turkey}
\address{$^{(2)}$ NIST  Center for Neutron Research,
National Institute of Standards and Technology, Gaithersburg, MD 20899 }
\address{$^{(3)}$ Department of Materials Science,
University of Pennsylvania, Philadelphia, PA 19104}
\date{March 9, 2000}
\maketitle
\preprint{To be appear in PRB}
\begin{abstract}
The band gap of a semiconducting single wall carbon nanotube decreases 
and eventually vanishes leading to metalization as a result of increasing
radial deformation. This sets in a band offset between the undeformed
and deformed regions of a single nanotube. 
Based on the superlattice calculations, we show 
that these features can be exploited to realize various quantum well
structures on a single nanotube with variable and reversible electronic 
properties. These quantum structures and nanodevices 
incorporate mechanics and electronics.
\end{abstract}

\bigskip

Unusual properties of electrons in the quantum structures
which were realized by using semiconductor heterostructures
($A_nB_m$) have initiated several fundamental studies\cite{1}.
Owing to the band offsets of the 
semiconductor heterostructures, the energies of the band states of one 
semiconductor $B$ may fall into the band gap of the adjacent semiconductor 
$A$. According to the effective mass approximation (EMA), 
the height (depth) of the
conduction (valence) band edge of $A$ from that of $B$, 
$\Delta E_{C}$ ($\Delta E_{V}$), behaves as a 
potential barrier for electrons (holes). For example,  $m$ layers 
of $B$ between $n$ layers of two $A$'s form a quantum well yielding
confined electronic states. 
The depth of the well and the width of the barrier and well
(in terms of number of layers $n$ and $m$, respectively) 
are crucial parameters to monitor 
the resulting electronic properties. 
Multiple quantum well structures
(MQWS) or resonant tunneling double barrier structures (RTDBS) can be 
tailored from the combination of $A$ and $B$, and from various stacking 
sequence of $n$ and $m$.

Single wall carbon nanotubes\cite{2} (SWNT) can display metallic 
or semiconducting character depending on their chiralities and 
diameters.\cite{3,4} Similar to the 
aforementioned idea exploited extensively in crystals\cite{1},
quantum structures can also be produced in SWNT's\cite{5,6,7,8,9}. 
It has been experimentally demonstrated that a rectifying
behavior can be achieved by the junction of two different 
SWNT's\cite{6}. Furthermore, the transport measurements on the
ropes\cite{7} and individual nanotubes\cite{8} have indicated 
a resonant tunneling behavior.
Recently, the quantum dot behavior has been also observed\cite{9}.
Like the semiconductor heterostructures, a different electronic 
property requires each time the fabrication of a new device using SWNT 
junctions. 

In this work, we propose a practical and interesting alternative, and show
that various quantum structures can easily be 
realized on an individual SWNT, and their electronic properties can 
be {\it variably} and {\it reversibly} monitored.
We predict and use the feature that the band gap of a semiconducting 
SWNT can be modified by radial deformation\cite{10} as described in
Fig. 1(a). More importantly, if such a deformation is not uniform 
but has different strength at different zones, each zone displays different 
band gap. Owing to the band offsets at the junction of different zones,  
MQWS or RTDBS of the desired electronic character 
can be formed, and novel electronic nanodevices can be engineered
on a single nanotube. This scheme is quite different 
from the previous constructions of SWNT heterostructures 
or quantum dots\cite{5}, where one had to fabricate each time a different 
junction or topological defects to satisfy the desired electronic 
character. 

First-principles calculations are carried out 
within the generalized gradient approximation (GGA) 
using plane waves with a cutoff energy of 500 eV and ultra 
soft pseudopotentials\cite{11,12}. Constrained structure optimizations 
are performed on the SWNT under transversal compression.
The zigzag (7,0) tube is a semiconductor when it is undeformed, but its 
band gap $E_g$ is modified upon introducing an elliptic deformation,
which is quantified by the ratio of the elliptic major axis to 
the minor axis $a/b$. The band 
gap decreases with increasing $a/b$, and eventually 
vanishes for $a/b \sim$ 1.28 (see Fig. 1(b)). After the onset of 
metalization upon the closure of the band gap, the density of states at 
the Fermi energy $D(E_F)$ increases with further increase of $a/b$,
as shown in Fig. 1(c). Relatively stronger radial deformations are
required for the closure of the gap of (8,0) and (9,0) tubes. The radial
deformation $a/b$, the strain energy per atom $E_s$, and the compressive 
force on the fixed atom $F_c$ at the closure of the band gap are calculated
to be ($a/b=1.28$; $E_s=18$ meV; $F_c=0.32$ eV/A), ($a/b=1.54$; $E_s=38$ 
meV; $F_c=0.31$ eV/A), and ($a/b=1.36$; $E_s=18$ meV; $F_c=0.25$ eV/A) for
(7,0), (8,0) and (9,0) tubes, respectively\cite{12}. The 
armchair (6,6) SWNT maintains the metallic behavior despite the 
radial deformation. Remarkably, the induced deformations, in particular
those causing to insulator-metal transition, are elastic. Our results confirm
that the atomic structure and hence the electronic properties return
to the original, undeformed state when the compressive stress is lifted.

Normally, as the radius $R \rightarrow \infty$ the electronic 
structure of a SWNT near the band edges becomes similar to that obtained 
by folding the $\pi^{*}-$ and $\pi-$bands of the graphene.
However, a singlet state at the 
band edge of a ($n$,0) SWNT with small $R$ involves significant 
$\sigma^{*}-\pi^{*}$ hybridization\cite{3}. This state occurs 
above the $\pi^{*}-$band ($n = 9$), but falls in the gap ($n = 7, 8$)
and eventually closes the gap ($n = 6$) as $n$ decreases\cite{3,12}.
Apparently, the estimate of $E_{g}$ based on the simple extrapolation
using the experimental gaps of SWNT with relatively larger radius\cite{4}
is not valid for the (7,0) tube. In the present case, 
introducing radial deformation and hence increasing the
curvature at both ends of the elliptic major axis reduces the 
band gap of the (7,0) SWNT owing to the enhanced $\sigma^{*}-\pi^{*}$ 
hybridization.

We performe also TB total energy and electronic structure
calculations for the undeformed and uniformly deformed (7,0) SWNT by 
using  transferable parameters\cite{13} related to carbon 2$s$ and
2$p_{x,y,z}$ orbitals. The radial deformation has been induced by approaching 
two atoms of the unit cell at one end of the diameter to two similar 
atoms at the other end indicated by dark atoms in Fig. 1(a).
Once the transversal strain is set by fixing these four atoms, the rest of 
atoms are relaxed by the conjugate gradient method. 
We consider undeformed ($a/b=$1) and five different degrees of radial
deformation ($a/b>$1). 
The variation of $E_g$, first and second states of the conduction band, 
($c_1$ and $c_2$) and those of the valence band, ($v_1$ and $v_2$) 
are illustrated in Fig. 1(d) and 1(e). Data points on the curves shown 
in these figures correspond to different degrees of elliptic deformations 
from I to VI as described in Fig. 1(a). We note deviations between the
first-principles PW and empirical TB results perhaps due to 
the differences in the details of the deformations and limitations 
of the empirical method. It is also known that the band gap $E_g$ is 
usually underestimated by local density approximation (LDA) calculations.
On the other hand, as discussed before, the $\sigma^{*}-\pi^{*}$ 
hybridization effect is crucial for setting $E_{g}$, and first-principles
PW calculations described it better. While the present GGA calculations 
for the undeformed (7,0) tube finds $E_g = 0.242$ eV, an earlier 
LDA calculation\cite{3} predicted $E_{g} = 0.09$ eV.  Our TB calculations
using transferable parameters predict $E_g \sim 0.5$ eV and hence 
relatively stronger deformation is required to reduce the TB gap $E_g$ 
to 0.1 eV. Earlier TB calculations found $E_g \sim 1.0$ eV for undeformed
tube\cite{14}. 
Comparison of band gaps measured by STM spectroscopy\cite{6} with those
calculated by different methods, and an extensive analysis for their 
variation with the radial deformation applied on different SWNT's will be
presented elsewhere\cite{12}. Nevertheless, both methods (PW and TB)
predict here similar overall behavior for the
band gap variation with the radial deformation. We will use the  
TB method to study MQWS, since it allows us to treat a large number of 
carbon atoms, which cannot be treated easily with the present 
first-principles PW method. We will treat the (7,0) tube as a 
prototype system

Reducing $E_{g}$ and eventually the onset of insulator-metal transition,
and further metallization of certain SWNT's with increasing radial
deformation, and the reversible nature of all these sequence of physical
events incorporate important ingredients suitable to form novel
quantum structures and nanodevices.
We investigate MQWS as a generic system and demonstrate that one can 
generate electronic properties convenient for 
various device applications. To start, we consider a (7,0) zigzag 
nanotube, that is pressed to squash only at certain regions. 
We assume that the undeformed region $A$
of $n$ unit cells\cite{15}, and adjacent deformed region $B$ of $m$ unit 
cells
(one interface atomic layer at both side has intermediary deformation)
repeat periodically, so that the translational periodicity 
along the axis of the tube involves $n+m=$16 cells and 448 carbon atoms.
This tube forms a $(A_nB_m)$ superlattice of semiconductor 
heterostructure, where the band gap of $A$ is larger than that of $B$. 
Since $E_g(A; a/b=1) > E_g(B; a/b > 1)$, a band offset shall occur 
at the junction. The ($n=8$, $m=$8) supercells of
the superlattices are schematically described for two different
degrees of deformation in Fig. 2.
Note that at low degree of deformation the junction can be formed 
by using one interface layer,
while more interface layers may be necessary if $B$ is 
severely deformed or a graded junction is aimed.  

Experimental and theoretical
methods have been proposed in the past to determine the band offsets, and
hence to reveal the band diagram perpetuating along the superlattice axis.
For the present situation ambiguities exist in calculating alignments
of the band edges and to determine band diagram in real space 
including band bowing
due to the charge transfer between $A$ and $B$. Even if the band diagram
were known, it is not obvious whether EMA
is applicable for an individual, non uniformly deformed SWNT. Therefore,
instead of applying EMA to the 1D band diagram,
we directly calculate the electronic structure of the ($A_nB_m$)
superlattice on an individual (7,0) SWNT. The band alignment 
is not explicit, but it is indigenous to the method and hence the confined 
states shall be obtained directly from 
the present TB superlattice calculations\cite{16}. 

We performed calculations on three different superlattices 
described in Fig.~2(a), i.e.
$(A_8B_8)$; $(A_4B_{12})$; $(A_{12}B_4)$, and calculated the electronic 
states $\Psi_{i,{\bf k}}({\bf r})$ with band energy $E_{i,{\bf k}}$. 
Here, $i$ and {\bf k} are band index and wave vector of the superlattice 
along its axis. Because of flat superlattice bands, we considered only
the $\Gamma -$point in the superlattice Brillouin zone.
Figure 3 illustrates the local (or cell) density of states, 
${\cal L}(E, j)= \sum_{i,{\bf k}}\int_{j}d{\bf r}
|\Psi_{i,{\bf k}}({\bf r})|^2 \delta (E- E_{i,{\bf k}})$ and the state
density
$|\Psi_{i,{\bf k}}(j)|^2 = \int_{j}d{\bf r}|\Psi_{i,{\bf k}}({\bf r})|^2$
both integrated at each cell $j$ in the supercell.
${\cal L}(E, j)$ with higher density near the band edges of $B$ 
(i.e. small gap region) is 
due to quantum well states and hence is consistent 
with the discussion presented at the beginning.
Second peak of ${\cal L}(E, j)$ in the conduction band occurs in $A$,
and becomes well separated from the first peak in the superlattice 
$(A_4B_{12})$.
The well known behavior of MQWS is apparent with the confinement
of states at the band edges. The first states at the band 
edges of $B$, i.e. $c_1$ and $v_1$ are confined in B
suggesting a normal band offset. The confinement of the second state 
in the valence band, $v_2$ is rather weak. On the other hand, the 
second state of the conduction band, $c_2$ is not confined in the well
of $B$, but is localized at the barrier of $A$. 
It appears that the energy of $c_2$ occurs above the well in $B$, 
 and $c_2$ cannot match with the next higher energy state of $B$. 
Similar to that observed in short periodicity AlGaAs 
superlattices,\cite{16} this situation demonstrates
that the description of the superlattice electronic structure in terms
of 1D multiple square well states obtained within the simple EMA 
can fail owing to the band structure effects.  
The confinement of $c_1$ and $v_1$ states increases
with $n$, i.e. with the length of the barrier region. This is an expected
result, since the longer barrier prevents the tunneling of these states 
 through $A$. 
Also the energy of $c_1$ raises with decreasing $m$.
This is a direct consequence of the uncertainty principle.

Figure 4 shows the MQWS behavior of the superlattice described in Fig. 2(b),
where $B$ is strongly deformed. The state density and local density of states
indicate that the confined states $c_1$ and $v_1$ display relatively higher
localization at the interface. This situation originates from the
interface atomic structure connecting $A$ to strongly deformed $B$.
Different band offsets can also be realized by setting up different level
of deformations at both $A$ and $B$. Again, depending on the
level of the deformation $E_g(B; a/b > 1)$ can even be zero that makes a
metal-semiconductor superlattice structure. Furthermore, from the junction
of two metalized SWNT's having different $D(E_F)$ one can form a metal-metal
superlattice.

In MQWS, the truly 1D states are normally propagating with the wave vector 
${\bf k}$, and form a band structure. The bands become flatter with
increasing $n$, and eventually the band picture breaks down and states
become totally localized in the quantum well (or in $B$). This way
the superlattice is expected to experience a Mott metal-insulator 
transition. Furthermore, a randomly deformed SWNT can be an interesting
system to investigate electron localization in 1D. The modulating
or $\delta-$ doping of a MQWS or QWS 
(also quantum dot) may exhibit interesting effects on the
transport properties\cite{16}.
It is interesting to note that the resonance condition of a RTDBS with 
$A_{n'}B_{m'}A_{n'}$ having contacts to metal reservoirs from both ends
shall be monitored by the deformation and size of $B$. 
Strain or pressure nanogauges or variable nanoresistors can be developed
based on the fact that the metalization and hence the conductance of a
(7,0) nanotube can be changed with the applied deformation.
Also a junction $A_{n'}B_{m'}$ with metallic $B$ is expected to show a 
rectifying behavior. We also note that a 3D grid of MQWS can
be constructed by periodic stacking of tubes where quantum wells occur
at crossing points. The electronic properties of this system can be
varied with the stacking sequence and applied pressure. Finally, we point 
out that the recent experimental work\cite{17} which showed that
the controlled local deformation can be achieved.

In conclusion, we showed that the electronic properties of a
semiconducting SWNT can be modified by introducing radial
deformation which can be used to produce interesting quantum structures 
and devices on a single tube, such 
as MQWS, RTDBS, rectifying junction, and variable nanoresistor with
continuously tunable electronic properties.

This work  was  partially supported by the National Science
Foundation under Grant No. INT97-31014 and
T\"{U}B\.{I}TAK under Grant No. TBAG-1668(197 T 116).


\begin{figure}
\centerline{\psfig{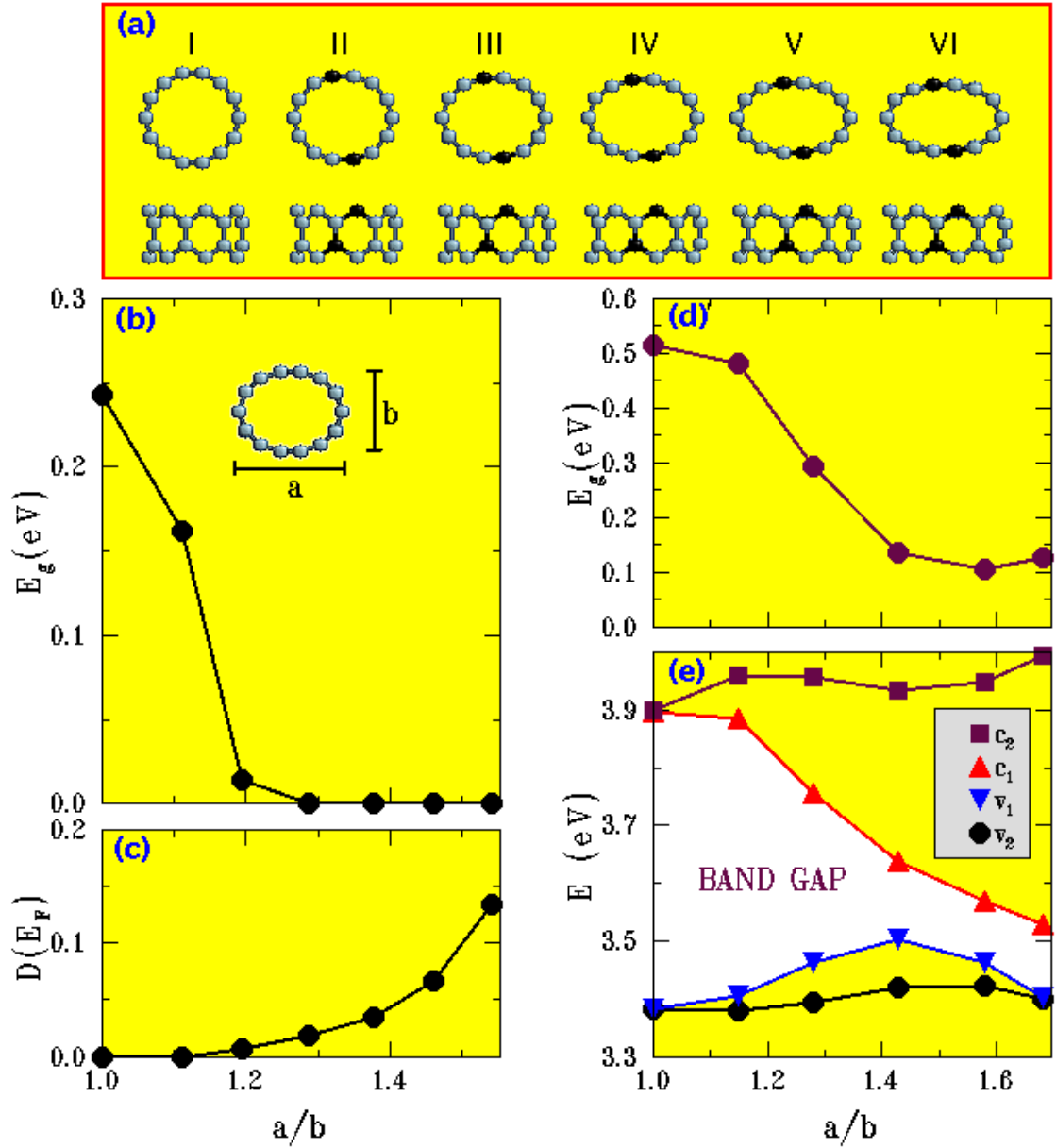}}
\vspace*{1cm}
\caption{(a) Top and side view of (primitive) unit cells of the (7,0) SWNT 
         under different degrees of elliptic (circumferential) deformation. 
         Undeformed
         tube with circular cross section is labeled by type-I. 
         (b) Variation of the energy
         band gap $E_g$, and (c) density of states 
         at the Fermi energy, $D(E_F)$
         with deformation $a/b$ calculated by the first-principles PW method.
         (d) Variation of $E_g$, and (e) first and second states 
         at the edge of the
         conduction band ($c_1$, $c_2$), 
         and first and second states at the edge of
         the valence band ($v_1$, $v_2$) with $a/b$ 
         calculated by the TB method.}
\end{figure}

\begin{figure}
\centerline{\psfig{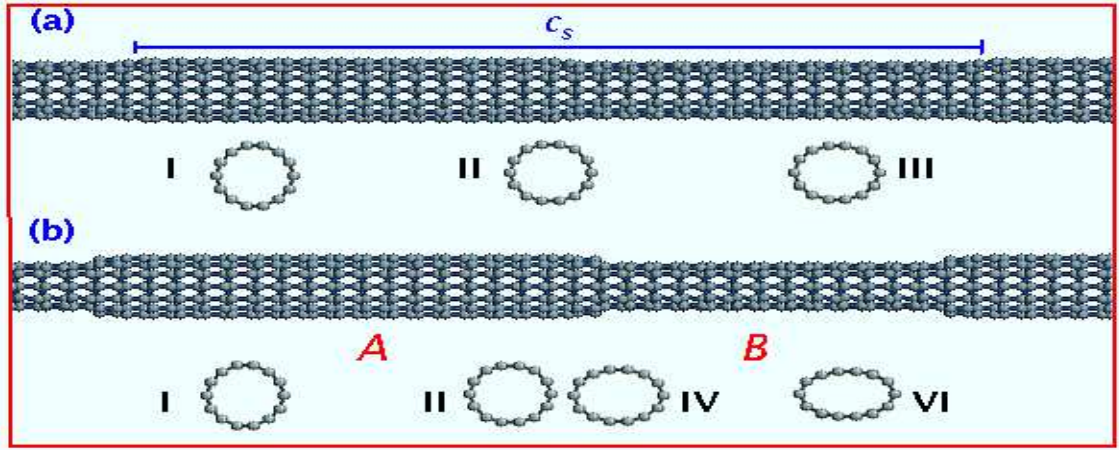}}
\vspace*{1cm}
\caption{Schematic descriptions of the ($A_8B_8$) supercells 
         of the superlattices 
         generated from the (7,0) SWNT. $c_s$ is the lattice parameter.
         (a) Region $A$ is undeformed, $B$
         has elliptic cross section of type-III. 
         The interface layer has type-II deformation. (b) Severely deformed 
         $B$ has type-VI deformation with two interface layers of type-II and
         IV deformation.}
\end{figure}

\begin{figure}
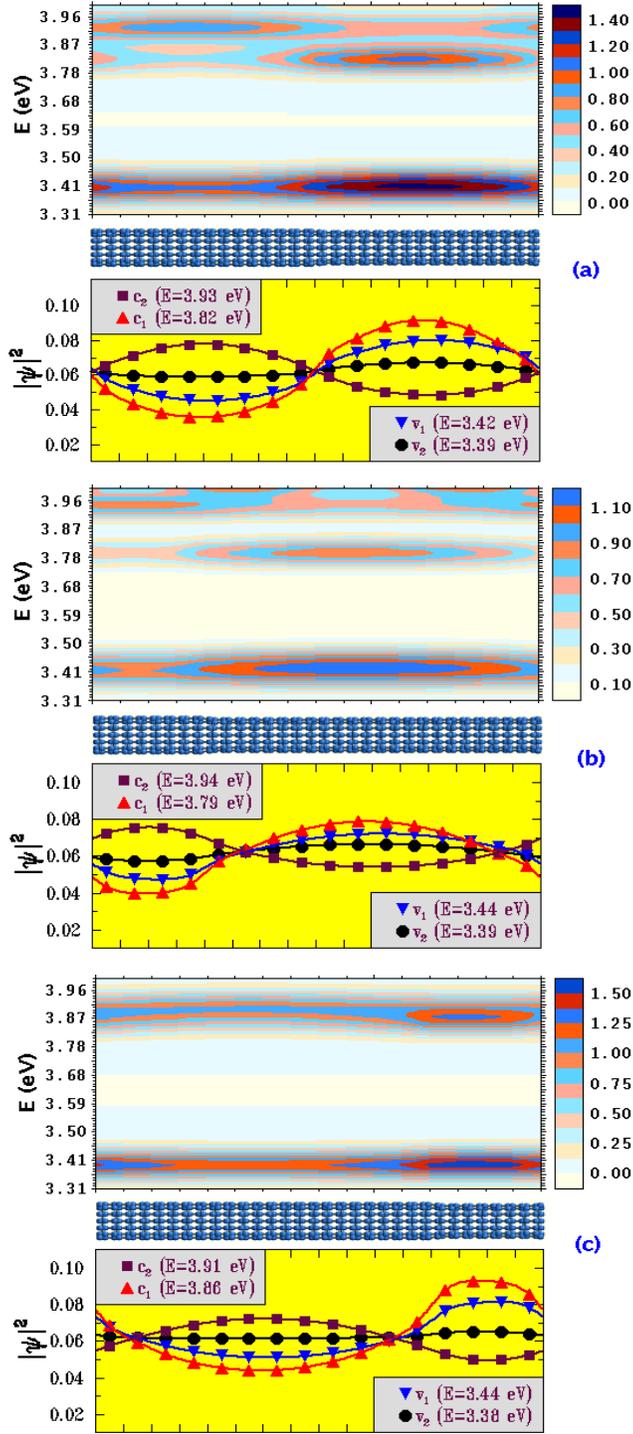

\centerline{\psfig{figure=fig3a.epsf,angle=0,width=85mm}}
\vspace*{0.2cm}
\centerline{\psfig{figure=fig3b.epsf,angle=0,width=85mm}}
\vspace*{0.2cm}
\centerline{\psfig{figure=fig3c.epsf,angle=0,width=85mm}}
\vspace*{1cm}
\caption{Upper panel: local density of states ${\cal L}(E,j)$; 
         lower panel: the state density $|\Psi_{i,{\bf k}}(j)|^2$ 
         at the cell $j$
         of the superlattices which display MQWS behavior. (a) $(A_8B_8)$; 
         (b) $(A_4B_{12})$; (c) $(A_{12}B_4)$.}
\end{figure}

\begin{figure}
\centerline{\psfig{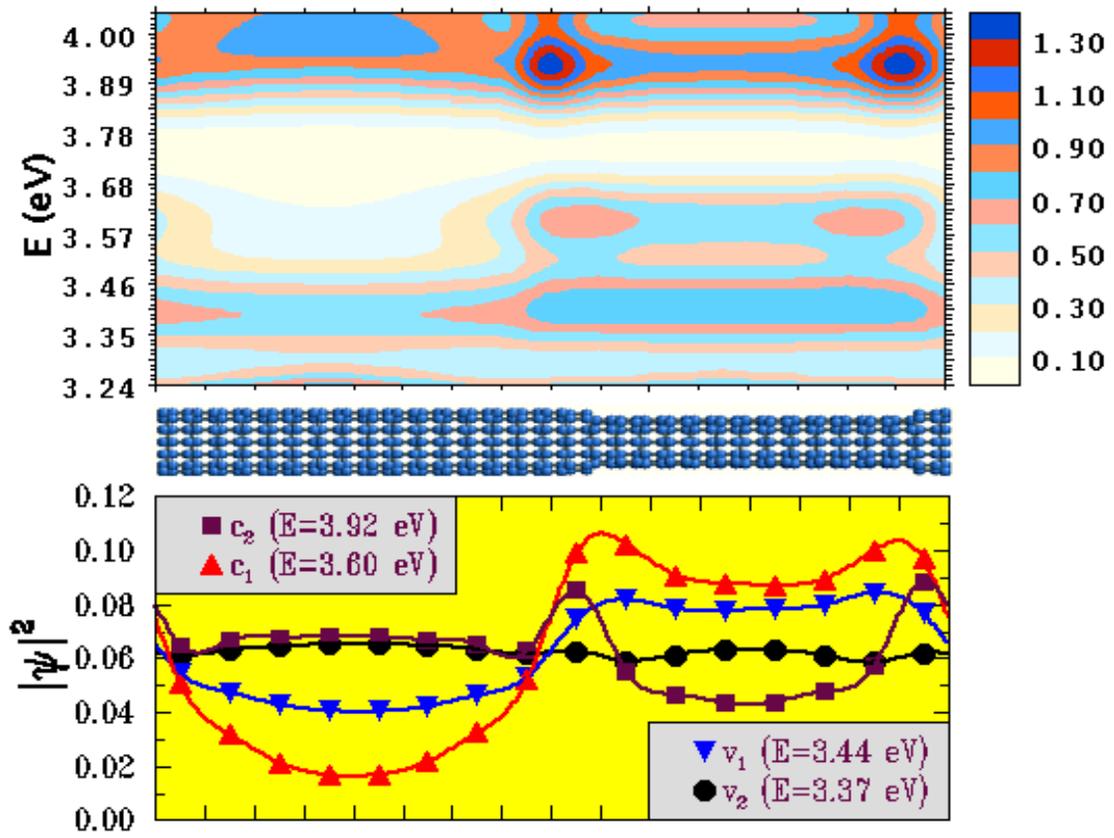}}
\vspace*{1cm}
\caption{The same as Fig.~3(a) except that $B$ has type-VI 
         cross section shown in Fig.~1(a).}
\end{figure}


\begin{references}
\bibitem{15} A cell denotes the primitive unit cell of the (7,0) nanotube,
             that has 28 carbon atoms. 
             The $A_8B_8$ supercell contains 16 such cells.
\end{references}
\end{document}